\documentclass[12pt,reqno]{article}
\usepackage{amsmath}
\usepackage{amsthm}
\usepackage{amsfonts}
\usepackage{amssymb}
\usepackage{epsfig}
\usepackage{graphics}
\usepackage{graphicx,indentfirst}
\numberwithin{equation}{section}
\usepackage{psfrag}
\usepackage{ccaption}

\def \Fig#1#2#3 {
\begin{figure}
\centering
\epsfxsize=#2cm \epsfbox{#1.eps}
\caption{#3}
\label{#1}
\end{figure}
}

\newcount\figno
\def\fig#1#2#3{
\par\begingroup\parindent=0pt\leftskip=1cm\rightskip=1cm\parindent=0pt
\baselineskip=15pt
\global\advance\figno by 1
\epsfxsize=#3
\centerline{\epsfbox{#2}}
\vskip 12pt
{\bf \small Figure \the\figno:} {\small #1}\par
\endgroup\par
}
\def\figlabel#1{\xdef#1{\the\figno
\mbox{ }}}
\def\encadremath#1{\vbox{\hrule\hbox{\vrule\kern8pt\vbox{\kern8pt
\hbox{$\displaystyle #1$}\kern8pt}
\kern8pt\vrule}\hrule}}

\newcommand{\Pm}{\Psi^-}
\newcommand{\Pp}{\Psi^+}
\newcommand{\gzero}{\mathfrak{g}_{\bar{0}}}
\newcommand{\gone}{\mathfrak{g}_{\bar{1}}}

\newcommand{\Om}{\Omega}

\newcommand{\p}{\partial_p}
\newcommand{\del}{\partial}
\newcommand{\Ad}{\text{Ad}}

\newcommand{\g}{\mathfrak{g}}
\newcommand{\m}{\mathfrak{M}}
\newcommand{\n}{\mathfrak{n}}

\newcommand{\h}{\mathfrak{h}}
\newcommand{\Z}{\mathbb{Z}}
\newcommand{\R}{\mathbb{R}}
\newcommand{\K}{\mathcal{K}}
\newcommand{\C}{\mathbb{C}}
\newcommand{\T}{\mathcal{T}}
\newcommand{\str}{\text{str}}
\newcommand{\tr}{\text{tr}}

\newcommand{\Out}{\text{Out}}
\newcommand{\cm}{c_-}
\newcommand{\cp}{c_+}
\newcommand{\cpm}{c_\pm}

\def\agl{{\rm $\widehat{\text{gl}}$(1$|$1)}}
\def\gl{{\rm gl(1$|$1)}}
\def\GL{{\rm GL(1$|$1)}}

\theoremstyle{break}

\newtheorem{thm}{Theorem}[section]
\newtheorem{prp}[thm]{Proposition}

\newtheorem{definition}[thm]{Definition}

\captionnamefont{\bfseries}
  \captiontitlefont{\small\sffamily}
  \captiondelim{ --- }
  \hangcaption

\author{Thomas Creutzig\\[5mm]
    DESY Theory Group, DESY Hamburg  \\
Notkestrasse 85, D-22607 Hamburg, Germany \\
\phantom{wwwx}{\small e-mail: }{\small\tt
thomas.creutzig@desy.de} }

\date{September 2008}

\begin{document}
\begin{titlepage}
    \title{\Huge Geometry of branes on supergroups}
    \maketitle       \thispagestyle{empty}

\vskip1cm
\begin{abstract}
In this note we analyze the geometry of maximally symmetric boundary conditions in Lie supergroup Wess-Zumino-Novikov-Witten models. We find that generically the worldvolume of a brane is a twisted superconjugacy class, very much like in the Lie group case. Whenever the brane is not completely delocalized in the fermionic directions a new atypical class of branes arises. We give an example of these new branes and show for type I supergroups and trivial gluing conditions that they can be naturally associated with atypical representations of the affine Lie superalgebra. 
\end{abstract}

\vspace*{-18.9cm}\noindent
{\tt {DESY 07-109}}\\
{\tt {arXiv:\,0708.0583}}

\end{titlepage}

\section{Introduction}

Conformal field theories on Lie supergroups and their cosets are an interesting area of current research. They have concrete applications in disordered electron systems and in string theory.  In addition sigma models with target space supersymmetry are interesting 
for their structural and mathematical properties, e.g. they provide examples of non-unitary models with possible nonpositive central charge. 

Studies of these models were initiated by Rozanski and Saleur \cite{Rozansky:1992td,Rozansky:1992zt}
who investigated the simplest non-trivial model the WZNW model on
the supergroup \GL.  These early investigations stimulated much further work on the
emerging topic of logarithmic conformal field theory (see e.g.\
\cite{Flohr:2001zs,Gaberdiel:2001tr} for a review). A few years
back, the \GL\ WZNW model was revisited in \cite{Schomerus:2005}
from a geometric rather than algebraic perspective. Effective computational 
tools were developed and then generalized to all type I Lie supergroups (\cite{Saleur:2006,Gotz:2006} and especially \cite{Quella:2007}).
The method was further developed to solve the boundary \GL\ model with volume filling branes \cite{Creutzig:2008}. 
Branes have been studied in the \GL\ model \cite{Creutzig:2007} and it was observed that the worldvolumes of the branes are twisted superconjugacy classes. 
\smallskip

The purpose of this note is two-fold. In the case of Lie groups the insight that the worldvolume of a brane is a twisted conjugacy class \cite{AS,St1,St2,Gawedzki,Felder:1999} turned out to be a powerfull aide in the analysis of the boundary WZNW model. 
Knowing the geometry of the brane is extremely helpfull in finding the gluing conditions of the fields, performing the minisuperspace analysis and setting up a formalism to solve the full quantum model. Also for problems like studying brane charges knowledge of the geometry is required. Thus for a further exploration of boundary WZNW models on Lie supergroups it is essential to understand its geometry. 

Representations of Lie superalgebras split into two classes, typical and atypical. While the appearance of singular vectors is generic in the representation theory of Lie algebras, singular vectors due to fermionic generators are quite atypical. Generalizing the proof that the branes' worldvolume is a twisted superconjugacy class from the known proof for Lie groups is generically possible but it fails at certain atypical points. These atypical points arise when the branes' worldvolume is not completely delocalized in the fermionic directions. We take this failure as a hint to suspect a different class of branes and indeed find them in the example of \GL. Precisely, associated to these atypical points, we find four possible gluing conditions of which one corresponds to a superconjugacy class, the others do not.
The failure is due to the fact that a certain operator is not invertible. This operator is not invertible because of the Grassmannian nature of the relevant Lie algebra. The boundary action typically requires the introduction of a 2-form $\omega$ which only exists if this operator is invertible. Thus for these new branes an action can only be found by introducing new auxiliary boundary fields, similar to \cite{Creutzig:2008}, which we can indeed do for \GL.   
Then we take a closer look at these branes and atypical representations. We restrict our attention to superconjugacy classes and representations of affine Lie superalgebras of type I. Superconjugacy classes can be labelled by an element of a chosen Cartan subalgebra. Since for basic Lie superalgebras there is a unique invariant non-degenerate bilinear form restricting non-degenerately to its Cartan subalgebras we can identify the Cartan subalgebra element with a weight and then with a representation associated to this weight. Via this identification, we show that a representation of the affine Lie superalgebra is atypical if and only if the associated superconjugacy class is localized in some fermionic directions.   
\smallskip

The structure of this note is as follows. 
In section 2 we recall some basic facts on Lie superalgebras and Lie supergroups. In section 3 we show that typically a branes' worldvolume is a twisted superconjugacy class. In section 4 we compute its action. In section 5 we investigate one example of atypical branes, then we compare co-adjoint orbits with representations of finite-dimensional Lie superalgebras of type I. This is the finite dimensional analogue of superconjugacy classes and representations of affine Lie superalgebras. Then we explain that if and only if a superconjugacy class is not completely delocalized in the fermionic directions the associated representation of the affine Lie superalgebra is atypical. In the appendix we list all gluing automorphisms of the relevant Lie superalgebras.

\section{Preliminaries}

We recall some properties of Lie superalgebras. The theory was developed by Victor Kac \cite{K1,K2}, a collection of results is given in \cite{FSS}. As a guideline to Lie supergroups we use the book by Berezin \cite{Berezin}. 

In the following, the Lie superalgebras will be over the real numbers $\R$. 
\begin{definition}\label{def:lsa}
Let $\g$ be a $\mathbb{Z}_2$ graded algebra $\g=\gzero\oplus \gone$ with product 
$[\ \ ,\ \ ]:\g\times\g\rightarrow\g$ that respects the grading. The parity of an homogeneous element is denoted by
\begin{equation}
	\begin{split}
|X|\ = \ \Bigl\{ \begin{array}{cc}
	       \ 0&\qquad X \ \ \text{in}\ \ \gzero    \\
	       \ 1&\qquad X \ \ \text{in}\ \ \gone     \\
                        \end{array} \ .
		\end{split}
\end{equation}
Then $\g$ is a {\bf Lie superalgebra} if it satisfies antisupersymmetry and graded Jacobi identity, i.e.
\begin{equation}
	\begin{split}
		0 \ &= \ [X,Y]+(-1)^{|X||Y|}[Y,X]  \qquad\text{and}\\[1mm]
	0\ &= \ (-1)^{|X||Z|}[X,[Y,Z]]+(-1)^{|Y||X|}[Y,[Z,X]]+(-1)^{|Z||Y|}[Z,[X,Y]]\ ,
\end{split}
\end{equation}
for all   $X,Y$ and $Z$ in $\mathfrak{g}$.

Further a bilinear form $B : \mathfrak{g} \times \mathfrak{g} \rightarrow \R$
is called a {\bf consistent supersymmetric invariant bilinear form} if
\begin{equation}
	\begin{split}
		B(X,Y) \ &= \ 0 \qquad\forall\, X \in \gzero \land \forall\, Y  \in \gone \\	
        	B(X,Y)-(-1)^{|X||Y|}B(Y,X ) \ &=\ 0 \qquad\forall\, X,Y\in \g \ \text{and}\\ 
	        B([X,Y],Z)-B(X,[Y,Z])\ &= \ 0 \qquad\forall\, X,Y,Z\in \g\ .\\  
\end{split}
\end{equation}	
\end{definition}

A simple Lie superalgebra whose even part is a reductive Lie algebra and which possesses a nonzero supersymmetric invariant bilinear form, is called
a basic Lie superalgebra. They are completely classified \cite{K1,K2}. There are the infinite series of unitary superalgebras
$sl(m|n)$ for $m\neq n$, $psl(n|n)$ and the orthosymplectic series $osp(m|2n)$ as well as some exceptional ones. In addition we will also consider Lie superalgebras of type $gl(m|n)$. We state their fundamental matrix realizations in appendix \ref{fundamentalreps}. A non degenerate invariant supersymmetric bilinear form is then given by the supertrace $\str$ in this matrix representation which we will use in the following sections.

A Lie supergroup can be obtained from a Lie superalgebra as follows. Let $\{t^a\}$ be a basis of $\gzero$ and $\{s^b\}$ a basis of $\gone$, then the Grassmann envelope $\Lambda(\g)$ of $\g$ consists of formal linear combinations 
\vspace{1mm}
\begin{equation}
	X \ = \ x_{a}t^a +\theta_{b}s^b \vspace{1mm}
\end{equation}
where the $x_a$ are Grassmann even, the $\theta_b$ Grassmann odd and summation over the indices is implied. Note, that $\Lambda(\g)$ is a Lie algebra. Then following Berezin \cite{Berezin} a supergroup $G$ is the group generated by elements $g$ of the form $g=\exp X$ with $X$ in the Grassmann envelope of $\g$, i.e. the Lie supergroup $G$ of the Lie superalgebra $\g$ is the Lie group of the Lie algebra $\Lambda(\g)$. Further we denote the Lie subgroup of the subalgebra $\gzero$ by $G_0$.


The Lie group $G$ acts on its Lie algebra $\Lambda(\mathfrak{g})$ by conjugation 
\vspace{1mm}
\begin{equation}
\Ad(a):\ \Lambda(\g) \ \rightarrow  \ \Lambda(\g)\ ,\qquad\qquad X\ \mapsto aXa^{-1}\vspace{1mm}
\end{equation}
for $a$ in $G$ and $X$ in $\Lambda(\g)$. Since the invariant bilinear form is the supertrace of a representation it is invariant 
under the adjoint action, i.e. \vspace{1mm}
\begin{equation}
\str(\Ad(a) X,\Ad(a) Y) \ =\ \str( X, Y)\vspace{1mm}
\end{equation}
for any $X,Y$ in $\Lambda(\g)$ and $a$ in $G$.

Consider a Lie supergroup $G$ with a supersymmetric invariant nonzero bilinear form. We identify the Grassmann envelope of the underlying Lie superalgebra with the tangent space at the
identity, $\Lambda(\mathfrak{g})=\T_e G$. 
On the tangent space $\T_g G$ at $g$ in $G$ we have left and right identification,\vspace{1mm}
\begin{equation}
\begin{split}
		&L_g\ :\ \Lambda(\mathfrak{g})\ \longrightarrow\ \T_gG, \qquad \ \ \ L_gX\ =\ gX \ \ \text{and}\\[1mm]
		&R_g\ :\ \Lambda(\mathfrak{g})\ \longrightarrow\ \T_gG, \qquad \ \ \ R_gX\ =\ Xg\ .\\[1mm]
	\end{split}
\end{equation}
The left identification induces a left invariant metric, i.e. $(gX,gY):=\str(X,Y)$. 
This metric is also right invariant, since it is invariant under the adjoint action $\Ad(g^{-1})$.

Now we turn to the description of branes. Let $\Sigma$ be an orientable Riemann surface with boundary $\del\Sigma$. Further let $x_a :\Sigma \rightarrow \Lambda_0(\R)$ and $\theta_b :\Sigma \rightarrow \Lambda_1(\R)$ be infinitely differentiable functions into the even respectively odd part of the Grassmann algebra over $\R$. By infinitely differentiable we mean a function of the form \cite{Berezin}\vspace{1mm}
\begin{equation}
	f \ = \ f(x) \ = \ \sum_{k\geq 0} \sum_{i_1,\dots,i_k}f_{i_1,\dots,i_k}(x)\theta_{i_1}\dots\theta_{i_k} \ ,\vspace{1mm}
\end{equation}
where $x\in\Sigma$, the $f_{i_1,\dots,i_k}(x)$ are $\R$-valued infinitely differentiable functions on $\Sigma$ and the $\theta_i$ generate the Grassmann algebra $\Lambda(\R)$. Locally one usually parametrizes any element in $\Sigma$ as $(\tau,\sigma)$, where the first coordinate belongs to the direction parallel to the boundary and the second one to the perpendicular direction, and then introduces complex variables $z=\tau+i\sigma$ and $\bar{z}=\tau-i\sigma$. Then we introduce the local $\Lambda(\g)$-valued field on $\Sigma$\vspace{1mm}
\begin{equation}
	X(z,\bar{z}) \ = \ x_a(z,\bar{z})t^a +\theta_b(z,\bar{z})s^b\vspace{1mm}
\end{equation}
and the $G$-valued field $g(z,\bar{z})=\exp X(z,\bar{z})$.
The currents of the WZNW theory at level $k$ are\vspace{1mm}
\begin{equation}
	J \ = \ -k\del gg^{-1} \qquad\text{and}\qquad \bar{J} \ = \ kg^{-1}\bar{\del}g\ . \vspace{1mm}
\end{equation}
Whenever the Lie superalgebra $\g$ is not simple there is not a unique invariant non-degenerate supersymmetric bilinear form. The holomorphic and anti-holomorphic Sugawara energy-momentum tensors $T$ and $\bar{T}$ are obtained by contracting the currents with the inverse of a distinguished invariant non-degenerate supersymmetric bilinear form (see \cite{Quella:2007} for the case of type I Lie superalgebras). The conformal symmetry of the WZNW theory is preserved if the energy-momentum tensor satisfies the boundary condition\vspace{1mm}
\begin{equation}
	T\ =\ \bar{T} \qquad\qquad \text{for} \ z\ =\ \bar{z}\ .\vspace{1mm}
\end{equation}
This is certainly satisfied if the boundary conditions of the currents are\vspace{1mm}
\begin{equation}\label{eq:gluing}
	J\ =\ \Omega(\bar{J}) \qquad\qquad \text{for} \ z\ =\ \bar{z}\ ,\vspace{1mm}
\end{equation}
where $\Om$ is an automorphism of $\g$ preserving any invariant non-degenerate supersymmetric bilinear form of $\g$.
The currents are $\Lambda(\g)$-valued fields and $\Om$ lifts to an automorphism of $\Lambda(\g)$ in the obvious way. Since these gluing conditions do not only preserve conformal symmetry but also half the current symmetry they are called maximally symmetric. 
Automorphisms of basic Lie superalgebras were classified by Vera Serganova \cite{Ser1}. 
We give a complete list of those that preserve the metric in appendix \ref{automorphisms}.

\section{Typical Branes}

For WZNW models on Lie groups the geometry of branes has been studied in detail e.g. \cite{AS,St1,Felder:1999}. 
If a field $g$ takes values in a Lie group with definite metric, then the boundary conditions (\ref{eq:gluing}) imply that the restriction of $g$ to the boundary of the Riemann surface $\Sigma$ takes values in a twisted conjugacy class. 

The generalization to Lie supergroups is the following. 
\begin{prp}
	Let the restriction of $g$ to the boundary of the Riemann surface $\Sigma$ take values in a subspace $N\subset G$ such that the boundary conditions (\ref{eq:gluing}) hold. We call $N$ the branes worldvolume. If the metric restricts non-degenerately to the tangent space $\T_gN$ of the branes worldvolume $N$ and if the tangent space of $G$ at the point $g$ decomposes in the direct sum of $\T_gN$ and its orthogonal complement $\T_gN^\perp$,\vspace{1mm}
	\begin{equation}\label{eq:orthogonal}
	\T_gG\ =\ \T_gN\oplus \T_gN^\perp\ ,\vspace{1mm}
\end{equation}
then the worldvolume $N$ is the twisted superconjugacy class\footnote{The automorphism $\Omega$ of the Lie superalgebra lifts to an automorphism of the Lie supergroup via $\Omega(\exp X)=\exp\Omega(X)$. We still denote it by $\Omega$.} \vspace{1mm}
\begin{equation} 
	C^\Omega_g\ = \ \{\ \Omega(h)gh^{-1} \ | \ h \ \in \ G \ \}\ .\vspace{1mm}
\end{equation}
\end{prp}
Since the metric is not definite, the decomposition \eqref{eq:orthogonal} is not guaranteed to hold in general. 
But for Lie supergroups with the property that the restriction of the metric to any simple or abelian subgroup of the underlying Lie group $G_0$ is definite 
it holds for a twisted superconjugacy class that is completely delocalized in the fermionic directions, 
i.e. $\exp\Lambda(\gone)\subset C^\Omega_g$. 
This is the regular case and we call these branes typical in analogy to typical representations. 
In section \ref{section:oddbranes} we will explain that this is more than a mere analogy.
We will call all other branes atypical. If the gluing automorphism $\Omega$ is inner, then the above assumptions also hold 
for non-regular twisted superconjugacy classes containing a point $g$ in the bosonic Lie subgroup $G_0$ while 
they never hold for twisted superconjugacy classes containing a point $g=\exp X$ with $X$ nilpotent. 
We will give an example in section \ref{section:GL} of branes covering these regions.
\smallskip

Now, let us explain the above proposition following \cite{St1}.
The gluing conditions  (\ref{eq:gluing}) can be translated to boundary conditions in the tangent space $\T_g G$ tangent to the point $g\in G$, i.e.
with the help of the left and right translation
the boundary conditions read\vspace{1mm}
\begin{equation}
	\del g\ =\ -\widetilde{\Omega}_g\bar{\del}g \vspace{1mm}
\end{equation}
where $\widetilde{\Omega}_g$ is the map on the tangent space at $g$ defined as\vspace{1mm}
\begin{equation}
	\widetilde{\Omega}_g\ =\ R_g\circ\Omega\circ L_{g^{-1}}\ : \ \T_gG \rightarrow \T_gG\ .\vspace{1mm}
\end{equation}
Note that for any given tangent vector $V$ in $\T_gG$ we have $\widetilde{\Omega}_g(V)=\Omega(g^{-1}V)g$.
In terms of Dirichlet and Neumann derivatives ($2\p=\del+\bar{\del}$ and $2i\del_n=\del-\bar{\del}$) the boundary conditions are\vspace{1mm}
\begin{equation}\label{eq:gluing2}
	(1+\widetilde{\Omega}_g)\p g\ =\ -i(1-\widetilde{\Omega}_g)\del_n g\ .\vspace{1mm}
\end{equation}

We need the assumption that the metric restricts non-degenerately to $\T_gN$ and that the tangent space $\T_gG$ splits into a direct sum\vspace{1mm}
\begin{equation}
	\T_gG\ =\ \T_gN\oplus \T_gN^\perp\ ,\vspace{1mm}
\end{equation}
then equation (\ref{eq:gluing2}) identifies those vectors in $\T_gG$ which have nonzero $(1-\widetilde{\Omega}_g)$ 
eigenvalue as directions of Neumann boundary conditions, i.e. they are vectors tangent to the branes worldvolume $N$. 
Then $T_gN^\perp$ is spanned by the vectors having zero $(1-\widetilde{\Omega}_g)$ eigenvalue.

Let $V$ be
in $\T_gN^\perp$ then $\widetilde{\Om}_g(V)=V$ which is expressed in terms of $\Om$\vspace{1mm} 
\begin{equation}
	\Om(g^{-1}V)\ =\ Vg^{-1}\ .\vspace{1mm}
\end{equation}
Therefore\vspace{1mm}
\begin{equation}\label{eq:step2}
	(\Om(g^{-1}V)-Vg^{-1},\Om(X))\ =\ 0\qquad \qquad \text{for all} \ X \ \text{in} \ \Lambda(\g)\ .\vspace{1mm}
\end{equation}
Since the metric is left- and right-invariant and invariant under $\Om$ (recall that $\Om$ is required to be metric preserving) \eqref{eq:step2} 
is equivalent to\vspace{1mm}
\begin{equation}
	(V,gX-\Om(X)g)\ =\ 0\ .\vspace{1mm}
\end{equation}
Hence $gX-\Om(X)g$ is orthogonal to $\T_gN^\perp$, i.e. it is tangent to the worldvolume $N$ of the brane,
but it is also tangent to the twisted superconjugacy class\vspace{1mm}
\begin{equation}
	C^{\Om}_g\ =\ \{\ \Om(h)gh^{-1}\ | \ h \ \in  G\ \}\ .\vspace{1mm}
\end{equation}
This can be seen as follows. Consider the curve $\Om(h(t))gh(t)^{-1}$ in $C^\Om_g$ through $g$, i.e. $h(0)=1$ and $\dot{h}(0)=-X$, then its tangent vector at $g$ is\vspace{1mm}
\begin{equation}
	\frac{d}{dt}\ \Om(h(t)) g h(t)^{-1}\Big|_{t=0}\ =\ gX-\Om(X)g\ .\vspace{1mm}
\end{equation}
Hence the tangent vectors of the form $gX-\Om(X)g$ are the tangent vectors of the twisted superconjugacy class $C^{\Om}_g$.
It remains to show that any tangent vector tangent to the worldvolume of the brane has the form $gX-\Om(X)g$.
Recall that those tangent vectors $V$ describe Dirichlet boundary conditions which are in the kernel
of $1-\widetilde{\Omega}_g$.
Therefore, the image of the adjoint operator $(1-\widetilde{\Om}_g)^\dagger$ must be $\T_gN$. 
Since $\widetilde{\Omega}_g=R_g\circ\Omega\circ L_{g^{-1}}$ is an isometry the adjoint is 
the inverse\vspace{1mm}
\begin{equation}
	(1-\widetilde{\Om}_g)^\dagger\ =\ (1-R_g\circ\Omega\circ L_{g^{-1}})^\dagger\ =\ 1-L_g\circ\Om^{-1}\circ R_{g^{-1}} \ ,\vspace{1mm}
\end{equation}
i.e. any element $W$ in $\T_gN$ can be written as \vspace{1mm}
\begin{equation}
	W\ =\ U-g\Om^{-1}(Ug^{-1})\vspace{1mm}
\end{equation}
for some $U$ in $\T_gG$. Further any vector $U$ in $\T_gG$ can be written as $U=\Om(X)g$ for some $X$ in $\Lambda(\g)$, hence
$W=\Om(X)g-gX$ for some $X$. We conclude that the worldvolume of a brane is a twisted superconjugacy class.
\smallskip

There are some remarks. 

Remark 1. The Lie supergroup acts on a twisted superconjugacy class by the twisted adjoint action $\Ad^\Om$\vspace{1mm}
\begin{equation}
         \begin{split}
              &\Ad^\Om(a): C^\Om_g\ \longrightarrow\ C^\Om_g \\[1mm]
              &\Ad^\Om(a)(\Om(h)gh^{-1})\ = \ \Om(a)\Om(h)gh^{-1}a^{-1} \ = \ \Om(ah)g(ah)^{-1} \\[1mm]
         \end{split}
\end{equation}
for any $a$ in $G$ and $\Om(h)gh^{-1}$ in $C^\Om_g$ . When analyzing branes on a Lie supergroup one usually starts with 
its semiclassical limit, the minisuperspace \cite{Creutzig:2008,Quella:2007b}.
The minisuperspace of a brane of a Lie supergroup is the space of functions that do not vanish on the brane. 
The infinitesimal twisted adjoint action acts on this space. This action is the semiclassical limit of the action of 
the boundary currents on the boundary fields.
The infinitesimal twisted adjoint action can be expressed through the infinitesimal lefttranslation 
(which is computed for type I Lie superalgebras in \cite{Quella:2007}). 
Let $h$ be in $G$ and $L_X^h$ be the lefttranslation in the direction $X$, i.e.\vspace{1mm}
\begin{equation}
    L_X^h: G\ \longrightarrow\ T_hG\ , \qquad L_X^hh\ =\ -Xh\ .\vspace{1mm}
\end{equation}  
Further its action on $h^{-1}$ is $L_X^hh^{-1}=h^{-1}X$ since $L_X(hh^{-1})=0$, hence its action on the twisted superconjugacy class element $a=\Om(h)gh^{-1}$ is
the infinitesimal twisted adjoint action\vspace{1mm}
\begin{equation}
         L_X^h(\Om(h)gh^{-1})\ = \ -\Om(Xh)gh^{-1}+\Om(h)gh^{-1}X \ = \ -\Om(X)a+aX\ .\vspace{1mm}
\end{equation}  
\smallskip

Remark 2. The stabilizer of $g$ under the twisted adjoint action is the twisted supercentralizer \vspace{1mm}
\begin{equation}
	\mathcal{Z}(g,\Om) \ = \ \{ h \ \in \ G \ | \ \Om(h)g\ = \ gh \ \}\ .\vspace{1mm}
\end{equation}
Its tangent space at $g$ is the kernel of $1-\tilde{\Om}_g$. The twisted superconjugacy class can be described by the homogeneous space $G/\mathcal{Z}(g,\Om)$. 
In the regular case the twisted supercentralizer is isomorphic to the maximal set $T^\Om$ of commuting points which are pointwise 
fixed under the action of $\Om$, i.e. it is contained in a maximal torus. Whenever $\Om$ is inner it is a maximal torus. 
A maximal torus of a basic Lie superalgebra is isomorphic to the maximal torus of its bosonic Lie subalgebra.
Therefore in the regular case the brane is completely delocalized  in the fermionic directions and since the metric is consistent (see Definition\ \ref{def:lsa})
the assumed orthogonal decomposition 
\eqref{eq:orthogonal} 
is true if it is true for the restriction to the Lie subgroup $G_0$.
At non regular points the brane is not necessarily completely delocalized in the fermionic directions in these cases one has to check wether \eqref{eq:orthogonal} holds.

It certainly does not hold for superconjugacy classes containing a point $g=\exp X$ with $X$ nilpotent, since then the operator $1-\tilde{\Om}_g$ is not diagonalizable. In this case there is a new type of branes whose geometry is rather different, we will give an example in section \ref{section:GL}.  
\smallskip

Remark 3. Gluing automorphisms must be metric preserving automorphisms of the relevant Lie algebra that is the Grassmann envelope $\Lambda(\g)$ of the Lie superalgebra $\g$. 
So far we obtained such an automorphism by lifting it from an automorphism of the Lie superalgebra $\g$. These are not all possible gluing automorphisms because conjugating by a fermionic Lie supergroup element $\theta$ is an automorphism of $\Lambda(\g)$ but not of $\g$. The above statements also hold for $\Om=\Ad(\theta)$ and a $\Ad(\theta)$ twisted superconjugacy class is simply a left translate by $\theta$ of an ordinary superconjugacy class.  

\section{The Action of typical branes}

In this section we state the action of WZNW models on supergroups. The Lie group case has been studied extensively and generalizes to Lie supergroups. In our considerations we will follow the reasoning of \cite{Gawedzki} and \cite{Mathieu}. We start with the bulk action.

\subsection{The bulk action}

The setup is exactly as in the Lie group case. So let $\Sigma$ be a compact Riemann surface without boundary, and $g:\Sigma\rightarrow G$ a map from the Riemann surface to the Lie supergroup $G$. Assume that there exists an extension of this map to a map $\tilde{g} : B\rightarrow G$ from a $3$-manifold $B$ with boundary $\del B=\Sigma$ to $G$. Further let $z=\tau+i\sigma$ and $\bar{z}=\tau-i\sigma$ then the kinetic term of the action is\vspace{1mm}
\begin{equation}
	S_{\text{kin}}[g] \ = \ \frac{k}{2\pi}\int_\Sigma d\tau d\sigma\ \str( g^{-1}\del g\ g^{-1}\bar{\del}g)  \vspace{1mm} 
\end{equation}
and the Wess-Zumino term is \cite{Witten}\vspace{1mm}
\begin{equation}
	S_{\text{WZ}}[\tilde{g}] \ = \ \frac{k}{2\pi}\int_B H \ = \ 
	\frac{k}{6\pi}\int_B \str(\tilde{g}^{-1}d \tilde{g}\wedge \tilde{g}^{-1}d\tilde{g}\wedge \tilde{g}^{-1}d \tilde{g})\ .  \vspace{1mm} 
\end{equation}
The full action is then \vspace{1mm}
\begin{equation}\label{eq:SWZW}
	S[\tilde{g}] \ = \  	S_{\text{kin}}[g] + S_{\text{WZ}}[\tilde{g}] \ .\vspace{1mm}
\end{equation}
Further the variation of the action is\vspace{1mm}
\begin{equation}
	\delta S \ = \ S[\tilde{g}+\delta\tilde{g}]-S[\tilde{g}]\ = \ 
	\frac{k}{\pi}\int_\Sigma d\tau d\sigma\ \str( g^{-1}\delta g\ \del(g^{-1}\bar{\del}g))\ . \vspace{1mm}
\end{equation}
Thus we obtain the bulk equations of motion $\del\bar{J}=\bar{\del}J=0$. 
As in the case of Lie groups it is straightforward to compute the Polyakov-Wiegmann identity\vspace{1mm}
\begin{equation}
	\begin{split}\label{PolyakovWiegmann}
		S[\tilde{g}\tilde{h}] \ = \  &S[\tilde{g}]+S[\tilde{h}]	+\frac{k}{\pi}\int_\Sigma d\tau d\sigma\ \str( \del hh^{-1}\ g^{-1}\bar{\del}g) \ .\vspace{1mm}
	\end{split}
\end{equation}
The action \eqref{eq:SWZW} is well-defined if it does not depend on the extension $\tilde{g}$ to a 3-manifold $B$.  
For type I Lie supergroup models this is done as follows \cite{Schomerus:2005,Gotz:2006,Quella:2007}. Type I Lie superalgebras have the distinguished $\Z$-graduation $\g=\g_-\oplus\g_0\oplus\g_+$, where $\g_\pm$ are two irreducible representations of the bosonic subgroup $\g_0$ and the supertrace satisfies\vspace{1mm}
\begin{equation}
	\str(X_+\,Y_+)\ = \ \str(X_-\,Y_-)\ = \ 0 \qquad \text{for all}\ X_\pm\ \text{and}\ Y_\pm \ \text{in} \ \g_\pm\ .\vspace{1mm}
\end{equation}
Then parameterizing a Lie supergroup element accordingly
$g=e^{\theta_-}g_0e^{\theta_+}$ and applying the Polyakov-Wiegmann identity \eqref{PolyakovWiegmann} twice the action becomes\vspace{1mm}
\begin{equation}
	S[\tilde{g}] \ = \ S[\tilde{g_0}] \ + \ \frac{k}{\pi}\int_\Sigma d\tau d\sigma\ \str( \Ad(g_0)(\del\theta_+)\bar{\del}\theta_-) \ .\vspace{1mm}
\end{equation}
Thus the ambiguity in the extension of this model is the ambiguity of the Lie group WZNW model of the bosonic subgroup $G_0$ and gives well-known quantization conditions to the level $k$ \cite{Gawedzki}.

Type II Lie supergroups can be treated similarly \cite{Hikida:2007}. The distinguished $\Z$-graduation is $\g=\g_{-2}\oplus\g_{-1}\oplus\g_0\oplus\g_1\oplus\g_2$ and the supertrace satisfies\vspace{1mm}
\begin{equation}
	\str(X_i\,Y_j)\ = \ 0 \qquad \text{for all}\ X_i\ \text{in}\ \g_i\ Y_j \ \text{in} \ \g_j\ \text{and} \ i+j\ \neq\ 0\ .\vspace{1mm}
\end{equation}
Then parameterizing a Lie supergroup element according to the distinguished $\Z$-graduation
$g=g_-e^{\theta_-}g_0e^{\theta_+}g_+$ and applying the Polyakov-Wiegmann identity \eqref{PolyakovWiegmann} four times the action becomes\vspace{1mm}
\begin{equation}
	\begin{split}
	S[\tilde{g}] \ = \ &S[\tilde{g_0}] \ + \ \frac{k}{\pi}\int_\Sigma d\tau d\sigma\ \str( \Ad(g_0)(\del\theta_+)\bar{\del}\theta_-)\ +\\
	&\frac{k}{2\pi}\int_\Sigma d\tau d\sigma\ \str( \Ad(g_0)([\theta_+,\del\theta_+]+2\del g_+g_+^{-1})
	([\bar{\del}\theta_-,\theta_-]+2g_-^{-1}\bar{\del} g_-)) \ .\vspace{1mm}
        \end{split}
\end{equation}
Thus the ambiguity in the extension in this model is the ambiguity of the Lie group WZNW model of the bosonic subgroup which corresponds to the Lie subalgebra $\g_0$. 

\subsection{The boundary action}

In this section we will state the boundary action. 
Following \cite{Gawedzki}, we represent a Riemann surface $\Sigma$ with boundary as $\Sigma'\backslash D$, where $\Sigma'$ is a Riemann surface without boundary and $D$ an open disc. We want to have a WZNW model based on a map $g:\Sigma\rightarrow G$ from the world-sheet with boundary to the Lie supergroup $G$. Therefore one needs to extend the map to a $3$-manifold $B$. This is not possible for a world-sheet with boundary. Thus the idea is to first extend $g$ to a map $g':\Sigma'\rightarrow G$ then to consider the WZ term based on $g'$ and to subtract a boundary term which only depends on the restriction of $g'$ to the closure of the disc $\bar{D}$.  
This boundary term has to be such that it coincides with the restriction of the Wess-Zumino term to the disc and such that the variation of the total action vanishes provided the usual equation of motion hold in the bulk and the desired gluing condition at the boundary.

Let us introduce the action and show that it has the two properties mentioned above. 

Let $g,g',\Sigma,\Sigma'$ as above, let $\tilde{g}:B\rightarrow G$ an extension of $g'$ to a 3-manifold $B$ with boundary $\del B=\Sigma'$. Further let the restriction of $g'$ to the closure of the disc $\bar{D}$ map to a twisted superconjugacy class $C^\Om_a$ at a regular point $a$,\vspace{1mm}
\begin{equation}
	g'(\bar{D})\ \subset\ C^\Om_a \ .\vspace{1mm}
\end{equation}
Then the WZNW action for the twisted boundary conditions $J=\Om(\bar{J})$ is given by\vspace{1mm}
\begin{equation}
	S_{\Omega,a}[g] \ = \  	S_{\text{kin}}^\Sigma[g] + S_{\text{WZ}}^B[\tilde{g}]-
	\frac{k}{2\pi}\int_{\bar{D}} \omega  \ .\vspace{1mm}
\end{equation}
Where $\omega$ is (using the shorthand $\tilde{\Omega}=\Ad(g'^{-1})\circ\Omega$)\vspace{1mm}
\begin{equation}
	\omega \ = \ \frac{1}{2}\str(g'^{-1}dg'\wedge \frac{\tilde{\Om}+1 }{\tilde{\Om}-1 }g'^{-1}dg')\vspace{1mm}
\end{equation}
and $\tilde{\Om}-1$ restricted to a twisted superconjugacy class is invertible as already seen in the previous section.
If we write $g'\vert_{\bar{D}}=\Omega(h)ah^{-1}$ then \vspace{1mm}
\begin{equation}
	(\tilde{\Om}-1)^{-1}g'^{-1}dg'\big\vert_{\bar{D}} \ = \ dhh^{-1} \ .\vspace{1mm}
\end{equation}
This allows us to rewrite the  boundary term as \vspace{1mm}
\begin{equation}\label{boundaryterm}
	\frac{k}{2\pi}\int_{\bar{D}} \omega \ = \ 
	\frac{k}{2\pi}\int_{\bar{D}} d\tau d\sigma \ \str(\tilde{\Om}(\del hh^{-1})\bar{\del}hh^{-1}-\tilde{\Om}(\bar{\del}hh^{-1})\del hh^{-1})\ .\vspace{1mm}
\end{equation}
Now we can check explicitly that the proposed action has the desired properties. First the restriction of the 3-form $H$ to the twisted superconjugacy class indeed coincides with $d\omega$\vspace{1mm}
\begin{equation}
	d\omega \ = \ H\big\vert_{C^\Om_a} \ .\vspace{1mm}
\end{equation}
Further the variation of the action vanishes provided the usual bulk equations of motion and the boundary equation of motions $J=\Omega(\bar{J})$ hold, \vspace{1mm}
\begin{equation}
	\begin{split}
		\delta S_{\Omega,a}[g]\ &= \ \delta S_{\Omega,a}^{\text{bulk}}[g]+
	\frac{ik}{2\pi}\int_{\del \bar{D}} d\tau \ 
	\str\bigl(\delta h h^{-1} \bigl((1-\tilde{\Omega})\bar{\del}hh^{-1}+(\tilde{\Omega}^{-1}-1)\del hh^{-1}\bigr)\bigr)\\[1mm]
	&= \  \delta S_{\Omega,a}^{\text{bulk}}[g]+\frac{i}{2\pi}\int_{\del \bar{D}} d\tau \ \str(\Omega(\delta h h^{-1}) (J-\Omega(\bar{J})))\ . \vspace{1mm}
        \end{split}
\end{equation}
A well-defined action should not depend on the extensions of the map $g$. 
In \cite{Creutzig:2008} the boundary GL($1|1$) model with gluing automorphism $\Omega=(-st)$ 
(see Appendix \ref{automorphisms} for the description of $(-st)$) is studied using a triangular decomposition of the group valued field. This can be generalized to all type I boundary WZNW models with gluing automorphism $\Omega=(-st)$ \cite{Creutzig:2008b} and as in the bulk case there is no additional quantization condition from the fermionic fields.

For general $\Om$ and any basic Lie superalgebra, there is a parameterization of the $G$-valued field $g$ that is particularly adapted to the problem: $g=\Omega(\theta)g_0\theta^{-1}$ where $g_0$ in the bosonic subgroup $G_0$ and $\theta$ takes values in $\exp\Lambda(\gone)$.
Using the Polyakov-Wiegmann identity \eqref{PolyakovWiegmann} and the explicit form of the boundary term \eqref{boundaryterm} one can rewrite the action as\vspace{1mm}
\begin{equation}
	\begin{split}
		S_{\Omega,a}[g] \ = \ S_{\Omega\vert_{G_0},a}[g_0] +
		\frac{k}{2\pi}\int_\Sigma d\tau d\sigma\
		&\str(\theta^{-1}\del\theta\ \theta^{-1}\bar{\del}\theta )+\str(\del g_0g_0^{-1}\ \Omega(\theta^{-1}\bar{\del}\theta) )+ \\[1mm]
		-\str(\theta^{-1}&\del\theta\ g_0^{-1}\bar{\del}g_0 )-\str(\theta^{-1}\del\theta\ g_0^{-1}\Omega(\theta^{-1}\bar{\del}\theta)g_0 ) \ . \\[1mm]
	\end{split}
\end{equation}
This model then has the same quantization conditions as the Lie group boundary WZNW model $S_{\Omega\vert_{G_0},a}[g_0]$.

\section{Atypical branes and atypical representations}\label{section:oddbranes}

The aim of this section is to investigate the existence of additional branes and its relation to atypical representations of finite dimensional Lie superalgebras and its affinizations. We start with an example of new additional branes.

\subsection{Atypical branes in the \GL\ WZNW model}\label{section:GL}

We restrict our attention for now to trivial gluing conditions $J=\bar{J}$.
We observed that there are certain points that do not intersect any typical brane with trivial boundary conditions. The natural question is whether there exist branes intersecting these points? We study this question in the example of \GL. The typical branes have been investigated in detail in \cite{Creutzig:2007}.  

The Lie superalgebra \gl\ is generated by two bosonic elements $E,N$ 
and two fermionic elements $\Psi^\pm$, subject to the relations\vspace{1mm} 
\begin{equation}
	[N,\Psi^\pm]\ =\ \pm\Psi^\pm \ \ , \ \ \{\Psi^-,\Psi^+\}\ =\ E \ \ \text{and} \ E \ \text{central}.\vspace{1mm}
\end{equation}
We parametrize a \GL\ valued field $g$ in terms of two bosonic fields $X$ and $Y$ 
and two fermionic fields $\cpm$, \vspace{1mm}
\begin{equation}\label{eq:par1}
	g\ =\ e^{i\cm\Pm}e^{iXE+iYN}e^{i\cp\Pp}\ .\vspace{1mm}
\end{equation}
Inserting our specific choice of the parametrization
\eqref{eq:par1}, the currents take the following form\vspace{1mm}
\begin{equation} \label{eq:J}
    \bar{J}\ = \ kie^{iY}\bar{\del}\cm\Pm + k\bigl(i\bar{\del}X -
     (\bar{\del}\cm)\cp e^{iY}\bigr) E + ki\bar{\del}Y N +
     k(i\bar{\del} \cp- \cp \bar{\del} Y)\Pp\vspace{1mm}
\end{equation}
and\vspace{1mm}
\begin{equation} \label{eq:bJ}
    J \ = \ -k(i\del\cm-\cm\del Y)\Pm -k\bigl(i\del X - \cm (\del \cp) e^{iY}\bigr)
    E -k i\del Y N -k ie^{iY}\del\cp\Pp.\vspace{1mm}
\end{equation}
In terms of these fields the gluing condition $J=\bar{J}$ imply the Dirichlet boundary condition $\del_\tau Y=0$ for the field $Y$. The non-regular points are those where $Y=y_0=2\pi s$, for an integer $s$, at the boundary. At these points the boundary conditions for the remaining fields are of the form\vspace{1mm}
\begin{equation}\label{glodd}
	2\del_\tau\cpm \ = \ \pm\cpm\del_\sigma Y \qquad \text{and} \qquad 2\del_\tau X\ + \ \cm\del_\sigma\cp+\cp\del_\sigma\cm \ =\ 0 .\vspace{1mm}
\end{equation}
Note that the first two equations imply $\del_\tau(\cm\cp)=0$. There are four different possibilities to satisfy these boundary conditions, i.e. by imposing the following Dirichlet conditions in addition to $Y = y_0 = 2\pi s$ for $s$ in $\Z$: 
  \newcounter{Lcount}
  \begin{list}{\Roman{Lcount}.}
    {\usecounter{Lcount}
    \setlength{\rightmargin}{\leftmargin}}
      \item \ \ $X =  x_0$ and $\cpm= 0$, 
      \item \ \ $\cp=0$,
      \item \ \ $\cm=0$ or 
      \item \ \ no additional ones. 
\end{list}
The first case corresponds to a superconjugacy class and has allready been investigated in \cite{Creutzig:2007} while the other cases are new and do not correspond to superconjugacy classes. 
The actions to these four cases requires the introduction of one extra bosonic boundary field $\beta$ and two fermionic boundary fields $\gamma_\pm$. 
Then the action is \vspace{1mm}
\begin{equation}
	S \ = \ \frac{k}{2\pi}\int d\tau d\sigma \ \del X\bar{\del}Y +\del Y\bar{\del}X+2e^{iY}\del\cp\bar{\del}\cm \ + \ S_{\text{bdy}}\vspace{1mm}
\end{equation}
where \vspace{1mm}
\begin{equation}
	\begin{split}
		S_{\text{bdy}}^{\text{I}} \ &= \ 0 \\[1mm]
		S_{\text{bdy}}^{\text{II}} \ &= \ \frac{k}{2\pi}\int_{\sigma=0} d\tau\ X\del_\tau\beta+\cm\del_\tau\gamma_+
		                             -\cm\gamma_+\del_\tau\beta \\[1mm]
		S_{\text{bdy}}^{\text{III}} \ &= \ \frac{k}{2\pi}\int_{\sigma=0} d\tau\ X\del_\tau\beta+\cp\del_\tau\gamma_-+\cp\gamma_-\del_\tau\beta \\[1mm] 
		S_{\text{bdy}}^{\text{IV}} \ &= \  \frac{k}{2\pi}\int_{\sigma=0} d\tau\ X\del_\tau\beta+\cm\del_\tau\gamma_+
		                    +\cp\del_\tau\gamma_--\cm\gamma_+\del_\tau\beta+\cp\gamma_-\del_\tau\beta \ . \vspace{1mm} 
	\end{split}
\end{equation}
The variation of the action under the corresponding Dirichlet conditions listed above vanishes along the boundary provided \eqref{glodd} holds. 

\begin{figure}[htb!]
	\label{fig:projectivecover}
\centering%
\psfrag{n}{$|0\rangle$}
\psfrag{n+}{$\psi^+|0\rangle$}
\psfrag{n-}{$\psi^-|0\rangle$}
\psfrag{n+-}{$\pm\psi^+\psi^-|0\rangle$}
\psfrag{p-}{$\psi^-$}
\psfrag{p+}{$\psi^+$}   
\includegraphics[width=6cm]{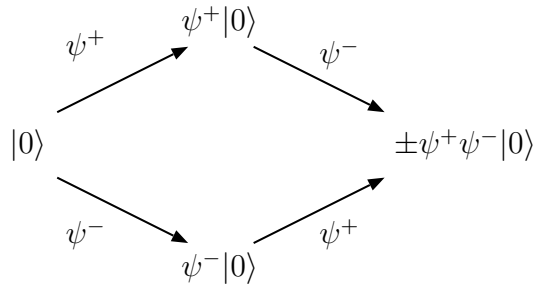}
\caption{{Projective cover $\mathcal{P}_0$: all states are annihilated by $E$ and $N$ and $\psi^\pm$ act as indicated. 
There is a 3-dimensional subrepresentation, two 2-dimensional ones and the trivial 1-dimensional subrepresentation.}}
\end{figure}

In order to get an idea what the spectrum of these new branes is let us look at the minisuperspace limit, that is the limit of large level $k$.
The minisuperspace limit of a boundary theory is given by the functions on the Lie supergroup modulo those that vanish on the brane. The case of typical branes has been analyzed in \cite{Creutzig:2007}.
The semi-classical spectrum of typical regular branes is the four dimensional adjoint representation $\mathcal{P}_0$. In the same way one can show that the minisuperspace limit of the above boundary conditions is built out of representations that are quotient of $\mathcal{P}_0$ by its proper invariant subrepresentations (there are four of them see Figure 1). In case I this is the trivial irreducible representation while in the other three cases the representations are not irreducible but they are still indecomposable. All these four representations are called atypical. 


\subsection{Geometry and irreducible Representations}

In this section all Lie superalgebras are basic simple Lie superalgebras of type I.

The co-adjoint orbit method of Kirilov and Kostant \cite{kirillov} relates co-adjoint orbits of a Lie group to representations of the group. In the case of compact simple Lie groups this correspondence is (\cite{Kirillovmerits} and references therein)\vspace{1mm}
\begin{equation}
	\pi_\lambda \ \longleftrightarrow \ \Omega_{\lambda+\rho}\vspace{1mm}
\end{equation}
where $\pi_\lambda$ is a irreducible highest weight representation of the compact Lie group $G$ with dominant highest weight $\lambda$, $\rho$ the Weyl vector and $\Omega$ the co-adjoint orbit in the dual of the Lie algebra $\g^*$ containing $\lambda+\rho$.  

Non compact Lie supergroups are relevant (i.e. real forms whose restriction to the bosonic subgroup is non compact), and the representations are neither necessarily finite nor highest(lowest) weight. 
In subsection \ref{section:finitereps} we relate representations of the finite dimensional Lie superalgebra $\g$ to its co-adjoint orbits,\vspace{1mm}
\begin{equation}
		\m_\lambda \ \longleftrightarrow \ \Omega_{\lambda+\rho} \ .\vspace{1mm}
\end{equation}
Here $\m_\lambda$ denotes a collection of representations associated to the weight $\lambda$. We show that the geometry encodes information about the representation in the following sense: if a representation $\pi_\lambda$ in $\m_\lambda$ is atypical then the associated co-adjoint orbit $\Omega_{\lambda+\rho}$ is not completely delocalized in the fermionic directions. Atypicality refers to the existence of fermionic singular vectors which we will explain below. In the previous section we saw in the example of \GL\ that these atypical representations are the minisuperspace limit of the spectrum of the new atypical branes.  

The next step in subsection \ref{section:affinereps} is to relate in a similar manner superconjugacy classes to representations of the affine Lie superalgebra at fixed level $k$ (we restrict our attention to highest(lowest)-weight representations), i.e.\vspace{1mm}
\begin{equation}
	V_\pm(\lambda,k) \ \longleftrightarrow \  (C_{\lambda+\rho},k) \ .\vspace{1mm}
\end{equation}
Here $V_\pm(\lambda,k)$ are highest(lowest)-weight representations and the correspondence is such that any atypical representation is associated to a superconjugacy class that is not completely delocalized in the fermionic directions.

In the case of the \GL\ model this correspondence has an interpretation in terms of Cardy boundary states.  

\subsubsection{Representations of finite dimensional Lie superalgebras}\label{section:finitereps}

The finite dimensional representations of finite dimensional classical Lie superalgebras are described by Kac in \cite{K1} and \cite{K3}.
Gould gives a generalization to infinite dimensional representations \cite{gould}.

Fix a Cartan subalgebra $\h$ of $\g$ and denote the dual space by $\h^*$. A non degenerate supersymmetric invariant bilinear form of the classical Lie superalgebras restricts non degenerately to a Cartan subalgebra $\h$ and induces a non degenerate bilinear form on its dual space. We denote it by $(\ \ | \ \ )$. Further a root is defined as follows.
\begin{definition}
	For $\alpha\neq0$ in $\h^*$ one sets
	\begin{equation}
		\g_\alpha \ = \ \{ a \ \in \ \g \ | \ [h,a]\ = \ \alpha(h)a \ \forall \ h \ \in \h \ \}\ .
	\end{equation}
	$\alpha$ is called a {\bf root} if $\g_\alpha\neq0$ and $\g_\alpha$ is called {\bf rootspace}. Further a root is called even if $\g_\alpha\cap\gzero\neq0$ and odd if $\g_\alpha\cap\gone\neq0$. Denote by $\Delta$ the set of roots, by $\Delta_0$ the set of even roots and by $\Delta_1$ the set of odd roots. 
\end{definition}
The Lie superalgebra $\g$ possesses the usual nonunique decomposition\vspace{1mm}
\begin{equation}
	\g\ = \ \n_- \oplus \h  \oplus \n_+ \ .\vspace{1mm}
\end{equation}
One calls a root positive if $\g_\alpha\cap\n_+\neq0$ and negative if $\g_\alpha\cap\n_-\neq0$. Let $\rho_0$ be half the sum of even positive roots and $\rho_1$ half the sum of odd positive roots, then the Weyl vector is \vspace{1mm}
\begin{equation}
	\rho\ =\ \rho_0-\rho_1\ .\vspace{1mm}
\end{equation}
We recall the classification results for irreducible representations of type I Lie superalgebras by Gould \cite{gould}. 
Let $\lambda$ in $\h^*$ be the highest weight of a highest weight representation $V(\lambda)$ and let $Z$ be the center of the universal enveloping algebra $U(\g)$ of the Lie superalgebra $\g$, then $Z$ takes constant values on $V(\lambda)$. The eigenvalue of $z$ in $Z$ on $V(\lambda)$ is denoted by $\chi_\lambda(z)$, this defines an algebra homomorphism\vspace{1mm}
\begin{equation}
	\chi_\lambda : Z\rightarrow \C \ ,  \qquad z\mapsto \chi_\lambda(z)\vspace{1mm}
\end{equation}
called infinitesimal character. A representation admits an infinitesimal character $\chi_\lambda$ if the elements $z$ in $Z$ take constant values 
$\chi_\lambda(z)$ in the representation. In the case of simple Lie algebras it is well known that every irreducible representation admits an infinitesimal character \cite{humphreys}. The generalization to type I Lie superalgebras is proved by Gould \cite{gould}:
\begin{thm}
	Every irreducible representation of a Lie superalgebra of type I admits an infinitesimal character $\chi_\lambda$ for some $\lambda$ in $\h^*$. 
\end{thm}
We construct representations explicitly as done by Kac \cite{K3}. 
Recall the triangular decomposition of type I Lie superalgebras $\g=\g_{-}\oplus\gzero\oplus\g_+$ ($\g_-$ and $\g_+$ are two irreducible representations of the bosonic subalgebra $\gzero$). Let $V_0$ be a representation of the bosonic subalgebra $\gzero$ of countable dimension then one gets a representation of $\gzero\oplus\g_+$ by promoting the elements in $\g_+$ to annihilation operators $\g_+(V_0)=0$ and the elements in $\g_-$ to creation operators, i.e. we define the Kac module of $V_0$ to be \vspace{1mm}
\begin{equation}
	\K(V_0)\ = \ \text{Ind}_{\gzero\oplus\g_+}^\g(V_0)\ .\vspace{1mm}
\end{equation}
The main results in \cite{gould} are summarized in
\begin{thm} Let $V_0$ be an irreducible representation of $\gzero$ and $\K(V_0)$ the Kac module. Then 
	\begin{itemize}
		\item there exists a maximal proper submodule $M(V_0)$, 
		\item the quotient $\K(V_0)/M(V_0)$ is irreducible and all irreducible representations are of this form. 
	        \item $V_0$ admits an infinitesimal character $\chi_\lambda^0$ and $\K(V_0)$ is irreducible if and 
			only if $(\lambda+\rho|\alpha)\neq0$ for all odd positive roots $\alpha$.
		\end{itemize}
\end{thm}
Denote the collection of Kac-modules with infinitesimal character $\chi_\lambda$ by $\m_\lambda$.  
In view of this theorem we call a representation $V_\lambda$ in $\m_\lambda$ atypical if there exists an odd positive root $\alpha$ such that $(\lambda+\rho|\alpha)=0$.

Now, we turn to co-adjoint orbits. Since the metric restricts non-degenerately to $\h$ there exists an $h_{\lambda+\rho}$ in $\h$ such that $(\lambda+\rho)(h)=(h_{\lambda+\rho},h)$ for all $h$ in $\h$. We write $\lambda+\rho=(h_{\lambda+\rho},\ \cdot \ )$, then the co-adjoint orbit containing $\lambda+\rho$ is\vspace{1mm}
\begin{equation}
	\Omega_{\lambda+\rho}\ = \ \{\ (gh_{\lambda+\rho}g^{-1}, \ \cdot\ ) \ | \ g \ \text{in} \ G \ \} \ .\vspace{1mm}
\end{equation}
It follows that the orbit extends into the dual space of the root space of the root $\alpha$ $\g^*_\alpha$ if and only if $(\lambda+\rho|\alpha)\neq0$.
This gives us the following relation between Kac-modules and co-adjoint orbits. 
\begin{prp}\label{coadjointfinite}
	There is a one-to-one correspondence between collections of Kac-modules with infinitesimal character $\chi_\lambda$ and co-adjoint orbits 
	\begin{equation}
		\Omega_{\lambda+\rho} \ \longleftrightarrow \ \m_\lambda \ ,
	\end{equation}
	such that if and only if a representation is atypical the associated co-adjoint orbit is not completely delocalized in the fermionic directions.
\end{prp}

\subsubsection{Representations of affine Lie superalgebras}\label{section:affinereps}

References to affine Lie superalgebras are \cite{Kac:1994} and \cite{Kac:2000}.
The affine Lie superalgebra corresponding to $\g$ is\vspace{1mm}
\begin{equation}
 \widehat{\g} \ =\  \C[t,t^{-1}]\otimes \g \oplus \C K \oplus \C d \vspace{1mm}
\end{equation}
where $K$ is central and\vspace{1mm}
\begin{equation}
	\begin{split}
[ \, t^m\otimes x , t^n\otimes y \, ]\ &=\ 
                    t^{m+n} \otimes [x,y] + m \delta_{m+n} (x,y) K   \\
[ \, d , t^n\otimes y \, ]\ &=\ n t^n\otimes y   \, .\vspace{1mm}
         \end{split}
\end{equation}
The vector space \vspace{1mm}
\begin{equation}
    \widehat{\h}\ =\ \h \oplus \C K \oplus \C d   \vspace{1mm}
\end{equation}
is a Cartan subalgebra of $\widehat{\g}$. We extend a linear function $\lambda$ on $\h$ to $\widehat{\h}$ by setting $\lambda(K) = \lambda(d) = 0$ and define linear functions $\Lambda_0$ and $\delta$ on $\widehat{\h}$ by\vspace{1mm}
\begin{equation}
	\Lambda_0 (\h \oplus \C d) \ =\  0 \ \ \ \ ,\ \ \ \ \Lambda_0 (K) \ =\  1 \ \ \ \ ,\ \ \ \
	\delta (h \oplus \C K)   \ =\ 0 \ \ \ \ \text{and}\ \ \ \ \delta(d)    \ = \ 1 \, .\vspace{1mm}
\end{equation}
Then $\widehat{\h}^* = \h^* \oplus \C \Lambda_0 \oplus \C \delta$.
We also extend a bilinear form $(\ \ ,\ \ )$ from $\g$ to $\widehat{\g}$ by setting\vspace{1mm}
\begin{equation}
	\begin{split}
		( t^m\otimes x , t^n\otimes y ) \ =\ \delta_{m+n} (x,y) \qquad &,\qquad
		( t^m\otimes x , K )\ =\ ( t^m\otimes x , d )\ = \ 0\ , \\
		( K , K )\ =\ ( d , d )\ =\ 0 \qquad&\text{and}\qquad(K,d)\ =\ 1\ .\vspace{1mm}
\end{split}
\end{equation}
Further the space of positive roots is\vspace{1mm}
\begin{equation}
	\widehat{\Delta}_+\ = \ \Delta_+\cup\{\alpha+n\delta|n>0\} \ .\vspace{1mm}
\end{equation}
The affine Weyl vector is\vspace{1mm}
\begin{equation}
	\widehat{\rho} \ = \ \rho + h^\vee\Lambda_0 \ ,\vspace{1mm}
\end{equation}
where $h^\vee$ is the dual Coxeter number that is the eigenvalue of the quadratic Casimir of $\g$ in the adjoint representation.

Representations of a WZNW model are constructed as follows. First one considers a Verma module $V_\pm(\Lambda)$ of the derived subalgebra $\widehat{g}'$ with highest(lowest)-weight vector $\Lambda=\lambda+k\Lambda_0$ where $\lambda$ is  a weight of $\g$. Then via the Sugawara construction (see \cite{Kac:4} for affine Lie algebras and \cite{Quella:2007} for the supercase) the Verma module extends to a representation of the semidirect sum $\widehat{\g}'+Vir$ in particular by identifying the Virasoro zero mode $L_0$ with the derivation $d$ it extends to a $\widehat{\g}$ module. The highest-weight vector then has conformal dimension ($L_0$ eigenvalue)\vspace{1mm}
\begin{equation}
	h_\Lambda \ = \ \frac{(\Lambda+2\widehat{\rho}\ |\ \Lambda)}{2(k+h^\vee)}\ .\vspace{1mm}
\end{equation}
We call such a Verma module typical if all its singular vectors are inherited from the bosonic subalgebra, otherwise it is called atypical. 
According to \cite{Quella:2007} a necessary condition for atypicality is \vspace{1mm}
\begin{equation}\label{eq:affineatypical1}
	h_{\Lambda-\alpha'} \ = \ h_\Lambda+n\vspace{1mm}
\end{equation}
where $\alpha=\alpha'+n\delta$ for some integer $n$ and an odd root $\alpha'$ of $\g$. If $\alpha$ is a positive odd root the highest-weight representation $V_+(\Lambda)$ can be atypical, and if $\alpha$ is a negative odd root the lowest-weight representation $V_-(\Lambda)$ can be atypical. Equation \eqref{eq:affineatypical1} can be rewritten as \vspace{1mm}
\begin{equation}\label{eq:affineatypical2}
	(\Lambda+\widehat{\rho}\ |\ \alpha) \ = \ 0\ .\vspace{1mm}
\end{equation}
In \cite{Kac:1979} it is shown that this is exactly the atypicality condition for basic affine Lie superalgebras of type I.

Atypical representations are closely related to atypical representations of the horizontal subalgebra. We know that $V_\pm(\Lambda)$ is atypical if there is a singular vector on the level of the horizontal subalgebra $\g$. Concatenating the representation with an automorphism of $\widehat{\g}$ gives an isomorphic representation that is also atypical. 
The affine Weyl group induces automorphisms of the affine Lie superalgebra $\widehat{\g}$. 
The affine Weyl group is the automorphism group on the root and coroot systems and hence induces an automorphism on the affine Lie superalgebra since this in return is uniquely defined via its roots, coroots and Cartan subalgebra.  Denote by $M$ the $\Z$ span of the coroots of $\g$ and define the translation $t_\alpha$ as ($\alpha$ in $M$)\vspace{1mm}
\begin{equation}
	t_\alpha(\lambda) \ = \ \lambda + \lambda(K)\alpha-((\lambda|\alpha)+\frac{1}{2}(\alpha|\alpha)\lambda(K))\delta \ .\vspace{1mm}
\end{equation}
We denote the group of translations $\{ t_\alpha \ | \ \alpha \ \text{in}\ M \}$ by $T_M$. Then the affine Weyl group is \cite{Kac:1994} ($W$ denotes the Weyl group of $\g$)\vspace{1mm}
\begin{equation}
	\widehat{W} \ = \ W \ltimes T_M \ .\vspace{1mm}
\end{equation}
The translation $t_\alpha$ induces an isomorphism $\tilde{t}_\alpha$ on $\widehat{\g}$ which acts explicitly as \vspace{1mm}
\begin{equation}
	\begin{split}
		\tilde{t}_\alpha:\ \ \qquad  h \ &\mapsto\ h+\alpha(h)K \qquad\qquad \text{for} \ h \ \text{in}\ \h \\[1mm]
		                     K \ &\mapsto\ K \\
				     d \ &\mapsto\ d-h_\alpha-\frac{1}{2}(\alpha|\alpha)K \\
				     g_\beta\otimes t^n \ &\mapsto\ g_\beta\otimes t^{n-(\alpha|\beta)} \qquad\qquad\text{for}\ g_\beta \ \text{in} \ \g_\beta \ .\\[1mm]
        \end{split}
\end{equation}
If one knows the characters of the representations of $\widehat{\g}$ then one can identify the representations obtained by an automorphism via (the $h_1,\dots h_r$ form an orthonormal basis of $\h$)\vspace{1mm}
\begin{equation}
	\begin{split}
	\chi_{\rho\circ\tilde{t}_\alpha}(q,z_i) \ &= \ 
	\tr_\rho(q^{\tilde{t}_\alpha(d)}\ z_1^{\tilde{t}_\alpha(h_1)}\ \dots\ z_r^{\tilde{t}_\alpha(h_r)}\ (-1)^F) \\[1mm]
	&= \ q^{-\frac{k}{2}(\alpha|\alpha)}\ z_1^{\alpha(h_1)k}\ \dots\ z_r^{\alpha(h_r)k}\ \chi_\rho(q, z_iq^{-\alpha(h_i)}) \ .\\[1mm]
\end{split}
\end{equation}
If every representation has a unique character then this identification is exact. In the cases of \agl\footnote{Even though \gl\ is not classical the above statements hold} \cite{Schomerus:2005}, $\widehat{\text{su}}(2|1)$ \cite{Saleur:2006} and $\widehat{\text{psu}}(1,1|2)$ \cite{Gotz:2006} all atypical representations could be obtained in this way from representations that have a singular vector on the level of the horizontal subalgebra $\g$.

We saw that the geometry of co-adjoint orbits provided information whether the associated representations are atypical or not. In a similar manner one can relate superconjugacy classes to representations of the affine Lie superalgebra $\widehat{\g}$. Choose an element $h_{\lambda+\rho}$ of the bosonic subalgebra $\gzero$ and choose a Cartan subalgebra $\h$ containing $h_{\lambda+\rho}$. Then we consider the superconjugacy class containing the point $\exp{\frac{2\pi ih_{\lambda+\rho}}{k+h^\vee}}$,\vspace{1mm}
\begin{equation}
	C_a \ = \ \{\ gag^{-1} \ | \ g \ \text{in} \ G \ \} \ , \qquad a \ = \ \exp{\frac{2\pi ih_{\lambda+\rho}}{k+h^\vee}} \ .\vspace{1mm}
\end{equation}	
The superconjugacy class is localized into a fermionic direction corresponding to an odd root $\alpha$ of $\g$ if and only if 
$\alpha(h_{\lambda+\rho})=n(k+h^\vee)$ for some $n$ in $\Z$. But this is equivalent to \vspace{1mm}
\begin{equation}
	(\lambda+k\Lambda_0+\widehat{\rho}\, |\, \alpha-n\delta) \ = \ (\lambda+\rho\, |\, \alpha)-n(k+h^\vee)(\Lambda_0\, |\, \delta) \ = \ 0 \ .\vspace{1mm}
\end{equation}
Thus we arrive at the affine analogue of proposition \ref{coadjointfinite}
\begin{prp}
	There is a one-to-one correspondence between Verma modules $V_\pm(\Lambda=\lambda+k\Lambda_0)$ and superconjugacy classes 
	\begin{equation}
		C_{\exp{\frac{2\pi ih_{\lambda+\rho}}{k+h^\vee}}} \ \longleftrightarrow \ V_\pm(\Lambda) \ ,
	\end{equation}
	such that a representation is atypical if and only if the associated superconjugacy class is not completely delocalized in the fermionic directions.
\end{prp}
\smallskip 

In the case of compact simple Lie groups the correspondence has a interpretation in terms of Cardy boundary states \cite{Cardy}. 
To each irreducible finite dimensional highest-weight representation of highest-weight $\Lambda=\lambda+k\Lambda_0$ 
there exists a boundary state $B(\Lambda)$ and in the semiclassical limit $k\rightarrow \infty$ 
this state becomes a distribution concentrated on the conjugacy class $C_a$ ($a= \exp{\frac{2\pi ih_{\lambda+\rho}}{k+h^\vee}}$)\cite{Alekseev:1999}.

For Lie supergroups boundary states have been studied only in the case of \GL\ \cite{Creutzig:2007} and there this correspondence is also true
for a special choice of the bilinear form\footnote{\gl\ is not simple thus the invariant bilinear form is not unique}.

\section{Conclusions}

In this note we have studied the geometry of maximally symmetric branes on Lie supergroups.

Following the reasoning for WZNW models on Lie groups \cite{AS,St1} 
we saw that typically branes are localized along twisted superconjugacy classes. 
This was expected from the previous analysis of branes in the \GL\ WZNW model \cite{Creutzig:2007}. 
The rather surprising result is the observation that there exist additional atypical branes.  
 
The argument that branes are twisted superconjugacy classes required that the brane intersects the bosonic Lie subgroup, hence there are regions in the Lie supergroup
which are not covered by any superconjugacy class. 
This was the first hint to suspect another kind of brane. 
Then we looked explicitly for new branes in the simplest Lie supergroup WZNW model that is the \GL\ model. 
Generically its superconjugacy classes are completely delocalized in the fermionic directions
but there also exists a discrete one-parameter family of point-like superconjugacy classes.
In this case we found four distinct choices of gluing conditions for the fields. 
The action of the new branes required the introduction of additional boundary fields similar to \cite{Creutzig:2008}. 
The semiclassical limit of these models are representations of the finite dimensional Lie superalgebra \gl.
These four representations have the same infinitesimal character $\chi_0$ and they are the four quotients of the representation $\mathcal{P}_0$ by its 
proper invariant submodules. $\mathcal{P}_0$ is indecomposable and maximal in the sense that it is not the quotient of a larger indecomposable representation.
The representation associated to the point-like superconjugacy class is irreducible while the remaining three representations are indecomposable but
not irreducible.

Then we studied the relation between the geometry of branes and the representations of finite dimensional and affine Lie superalgebras, focusing on atypicality.
In the case of Lie groups, untwisted branes (i.e. with trivial gluing automorphism) are in one-to-one correspondence to conjugacy classes which are in one-to-one correspondence to 
irreducible representations of the current algebra at fixed level $k$.
For Lie superalgebras irreducible representations are in one-to-one correspondence to superconjugacy classes which correspond to untwisted branes. Whenever the superconjugacy class
is not completely delocalized in the fermionic directions the associated representation is atypical and we saw in the example of \GL\ that additional
atypical branes appear which we associate to representations that are indecomposable but not irreducible.
\smallskip

There are several interesting possible extensions of this work. 

The relevance of atypical branes should be studied in more detail. 
In general it would be interesting to compute correlation functions of boundary WZNW models on Lie supergroups. 
For the bulk models a Kac-Wakimoto like formulation involving $bc$ ghost systems at central charge $c=-2$ has been introduced in \cite{Saleur:2006,Gotz:2006,Quella:2007}.  
This formalism allows the perturbative expansion of correlation functions in terms of correlation functions of the WZNW model of the bosonic Lie subgroup and of the $bc$ system. Due to the zero modes of the ghosts, only a finite number of terms of the perturbative expansion can contribute. 
In \cite{Creutzig:2008} a similar setup has been introduced for the \GL\ model. 
For those gluing conditions where the ghost field $b$ is identified with its antiholomorphic counterpart $\bar{b}$ only a finite number of terms of the perturbative expansion contributed. This works for the gluing automorphism $(-st)$ introduced in the appendix. We are currently working on a Kac-Wakimoto like formulation for type I Lie supergroup boundary WZNW models with gluing automorphism $(-st)$. For trivial boundary conditions one can use the boundary conditions of \cite{Creutzig:2006} for the $bc$ system. Then a derivative of the field $c$ is identified with the field $\bar{b}$ and the perturbative expansion does not terminate.  It would be also interesting to be able to solve models with other gluing automorphisms than $(-st)$.

For the \GL\ model it is shown \cite{Creutzig:2008} that correlation functions only involving atypical (bulk and also boundary) fields coincide with
the untwisted sector of the symplectic fermions \cite{Kausch:2000}. This method also applies to the atypical branes in \GL.
A generalization to type I Lie supergroup models with vanishing dual Coxeter number 
might also be possible.

Another interesting direction is the study of branes and the group of brane charges which in the case of compact Lie groups has a geometric interpretation
as the twisted K-group of the Lie group \cite{Witten:1998}.  
\bigskip
\bigskip 
\bigskip

\noindent {\bf Acknowledgments:} I wish to thank 
 Thomas Quella, David Ridout, Peter R\o nne and in particular 
Volker Schomerus for interesting discussions and 
comments on issues related to this work and on the manuscript.

\appendix
\section{Appendix}

\subsection{Fundamental representations}
\label{fundamentalreps}
In this section, we provide the fundamental matrix realizations of the superalgebras gl$(m|n)$, sl$(m|n)$, psl$(n|n)$ and osp$(m|2n)$, we follow \cite{FSS}.

gl$(m|n)$ is given by
\begin{equation}
	\begin{split}
		\text{gl}(n|m)=\Bigl\{ \left(\begin{array}{cc}A & B \\ C & D\\ \end{array}\right) \Bigr\},
	\end{split}
\end{equation}
where $A$ and $D$ are square matrices of size $n$ and $m$, $B$ is a $n\times m$ matrix and $C$ is a $m\times n$ matrix. The supertrace is defined via 
\begin{equation}
	\begin{split}
		\str  \left(\begin{array}{cc}A & B \\ C & D\\ \end{array}\right) =\tr A -\tr D \ .
	\end{split}
\end{equation}
Then, we have the unitary superalgebra 
\begin{equation}
	\begin{split}
		\text{sl}(n|m)=\bigl\{ X \ \in \text{gl}(n|m) \ | \ \str X =0\bigr\},
	\end{split}
\end{equation}
for $n\neq m$, and for $n=m$ the projective unitary superalgebra psl$(n|n)=$sl$(n|n)/\mathcal{I}$, where
$\mathcal{I}$ is the one dimensional ideal generated by $1_{2n}$. 
Further, we have the the orthosymplectic series
\begin{equation}
	\begin{split}
		\text{osp}(m|2n)=\bigl\{ X \ \in \text{gl}(m|2n) \ | \ X^{st}B_{m,n}+B_{m,n}X=0\bigr\},
	\end{split}
\end{equation}
where the supertranspose is
\begin{equation}
	\begin{split}
		\left(\begin{array}{cc}A & B \\ C & D\\ \end{array}\right)^{st}=\left(\begin{array}{cc}A^t & -C^t \\ B^t & D^t\\\end{array}\right)
	\end{split}
\end{equation}
and 
\begin{equation}
	\begin{split}
		B_{m,n}=\left(\begin{array}{cc}1_m & 0 \\ 0 & J_n\end{array}\right) \ ,\text{where} \ 
			J_n=\left(\begin{array}{cc}0 & 1_n \\ -1_n & 0\end{array}\right).
	\end{split}
\end{equation}

\subsection{Gluing automorphisms}\label{automorphisms}

We already showed in the second section that inner automorphisms preserve the metric, it remains to find all metric preserving outer automorphisms. For complex Lie superalgebras the groups of outer automorphisms are classified in \cite{Ser1}.
Denote the group of outer automorphisms of $G$ by $\Out \ G$, then $\Out \ $sl$(m|n)=\mathbb{Z}_2$, 
generated by $(-st):X\mapsto -X^{st}$. It is straightforward to verify, that this is indeed an automorphism and that it 
preserves the supertrace.

Further, $\Out \ $osp$(2m+1|2n)=1$, and  $\Out \ $osp$(2m|2n)=\mathbb{Z}_2$, generated by $\Ad J_{m,n}$, where $J_{m,n}$
in $gl(2m,2n)$, with $\det \ J_{m,n} = -1, J_{m,n}^2=1_{2m+2n}$ and $J_{m,n}B_{2m,n}J_{m,n}=B_{2m,n}$.
This leaves also the metric invariant by the same argument as for an inner automorphism.

Before we turn to psl$(n|n)$, $n\neq2$, we introduce the relevant automorphisms.
For $X=\bigl(\substack{A \ B \\ C\ D}\bigr)$, we have ($\lambda \in \mathbb{C}$)
\begin{equation*}
        \Pi(X)=\left(\begin{array}{cc} D & C \\ B & A\end{array}\right)\ , \ \ \ 
		\delta_\lambda(X)= \left(\begin{array}{cc} A & \lambda B \\ \lambda^{-1}C & D\end{array}\right)\ .
\end{equation*}
Note, that $\Pi$ changes the metric by an overall minus sign, so it is not a gluing automorphism. Since $\delta_\lambda$ acts only on
the fermionic part, it preserves the metric.
Then $\Out \ $psl$(n|n)$ is described by the exact sequence 
\begin{equation*}
	1 \rightarrow \{\delta_\lambda\}/\{\delta_\lambda|\lambda^n=1\}\rightarrow\Out\ \text{psl}(n|n)
	\rightarrow\mathbb{Z}_2\oplus\mathbb{Z}_2\rightarrow 1,
\end{equation*}
where $\mathbb{Z}_2\oplus\mathbb{Z}_2=\Pi\oplus (-st)$.
psl$(2|2)$ carries an action of $SL(2)$ induced by
\begin{equation*}
	\ \left[\left(\begin{array}{cc} a & b \\ c & -a\end{array}\right)\ ,\
		\left(\begin{array}{cc} A & B \\ C & D\end{array}\right)\ \right] \ = \ \ 
			\left(\begin{array}{cc} 0 & aB+bJ_1C^tJ_1^{-1} \\ -cJ_1B^tJ_1^{-1}-aC & 0\end{array}\right)\,
\end{equation*}
this defines an automorphism and since it leaves the bosonic subalgebra invariant it preserves the metric.
$\Out \ $psl$(2|2)$ is described by the exact sequence 
\begin{equation*}
	1 \rightarrow SL(2)/(-1_2)\rightarrow\Out\ \text{psl}(2|2)
	\rightarrow\mathbb{Z}_2\rightarrow 1,
\end{equation*}
where $\mathbb{Z}_2$ is generated by $\Pi$.

Further, the extension of the automorphisms to gl$(n|m)$ is straightforward and the dilatation of the fermionic elements becomes inner. 

Note, that the exceptional classical Lie superalgebras do not admit a metric preserving outer automorphism. 

\subsection{Real Forms}

Real forms of classical Lie superalgebras are classified in \cite{Ser2} and \cite{P}.
As in the case of simple Lie algebras this is done by classifying the involutive semimorphisms of the complex Lie superalgebras. 
A semimorphism $\phi$ of a complex Lie superalgebra $\g$ is a semilinear transformation such that 
\begin{equation}
	[\phi(X),\phi(Y)] \ = \ \phi([X,Y]) \qquad \text{for all} \ X,Y \ \text{in} \ \g \ .
\end{equation}
Then for every involutive semimorphism $\phi$ 
\begin{equation}
	\g^\phi \ = \ \{ X + \phi(X) \ | \ X \ \text{in} \ \g \ \} 
\end{equation}
is a real classical Lie superalgebra and these are all (Theorem 2.5 in \cite{P}). 

The relevant Lie algebra is the Grassmann envelope $\Lambda(\g)$ of the Lie superalgebra $\g$ and there is the following real form that is not coming from a real form of a Lie superalgebra. Define the superstar operation as \cite{FSS}
\begin{equation}
	(c\theta)^{\#} \ = \ \bar{c}\theta^{\#} \ , 
	\qquad \theta^{\#\#}\ = \ -\theta \ , \qquad (\theta_1\theta_2)^{\#} \ = \ \theta_1^{\#}\theta_2^{\#} \ 
\end{equation}
for any Grassmann numbers $\theta, \theta_i$ and any complex number $c$. 
Then concatenation of the superstar $\#$ with the automorphism $(-st)$ is a semimorphism of $\Lambda(\g)$ giving rise to a real form of $\Lambda(\g)$. 
\smallskip

Further any automorphism $\Om$ of the Lie algebra $\Lambda(\g)$ restricts to an automorphism of $\Lambda(\g)^\phi$ if and only if it leaves $\Lambda(\g)^\phi$ invariant that is $\Om$ and $\phi$ commute.

\end{document}